\documentclass[9pt,conference]{IEEEtran}
\IEEEoverridecommandlockouts

\usepackage{cite}
\usepackage{amsmath,amssymb,amsfonts}
\usepackage{graphicx}
\usepackage{textcomp}
\usepackage{amsmath}
\usepackage{booktabs} 
\usepackage{multirow}
\usepackage{hyperref}
\usepackage{stfloats}
\usepackage[table]{xcolor}  
\definecolor{bestgreen}{RGB}{220,245,220}
\usepackage{float}
\usepackage{url}            
\usepackage{booktabs}       
\usepackage{amsfonts}       
\usepackage{nicefrac}       
\usepackage{microtype}      
\usepackage{xcolor}         
\usepackage{algorithm}
\usepackage{algpseudocode}
\usepackage{amsmath}
\usepackage{tikz}

\usepackage{orcidlink}
\usepackage{comment}
\usepackage{amssymb}
\usepackage{braket}
\usepackage{bm}
\usepackage[font=footnotesize,skip=0pt]{caption}
\usepackage{xcolor}
\def\BibTeX{{\rm B\kern-.05em{\sc i\kern-.025em b}\kern-.08em
    T\kern-.1667em\lower.7ex\hbox{E}\kern-.125emX}}
\begin{document}
\bstctlcite{bstctl:etal, bstctl:nodash, bstctl:simpurl}

\title{SPATE: Spiking-Phase Adaptive Temporal Encoding for Quantum Machine Learning}

\author{\IEEEauthorblockN{Nouhaila Innan\textsuperscript{1,2}, Rachmad Vidya Wicaksana Putra\textsuperscript{1}, and Muhammad Shafique\textsuperscript{1,2}}
\IEEEauthorblockA{
\textsuperscript{1}eBRAIN Lab, Division of Engineering, New York University Abu Dhabi (NYUAD), Abu Dhabi, UAE\\
\textsuperscript{2}Center for Quantum and Topological Systems (CQTS), NYUAD Research Institute, NYUAD, Abu Dhabi, UAE\\
nouhaila.innan@nyu.edu, rachmad.putra@nyu.edu, muhammad.shafique@nyu.edu\\
}}

\maketitle

\begin{abstract}
Most quantum machine learning (QML) pipelines still rely on static encodings such as angle and amplitude maps, and this limits their ability to handle temporal information. To address this limitation, this paper uses spike-based data representation as an effective encoding mechanism that incorporates temporal structure into quantum feature preparation. Specifically, we propose \textit{Spiking-Phase Adaptive Temporal Encoding} (SPATE), a novel spike-driven temporal encoding method that converts real-valued tabular features into leaky integrate-and-fire spike trains and maps spike statistics to quantum rotations, augmented with a small set of temporal qubits through controlled phase operations. An encoding-centric evaluation protocol is also introduced to assess representation quality independently of the classifier, covering centered kernel-target alignment (CKTA), Fisher-style separability, inter/intra-class distance ratios, silhouette score, normalized entropy, and pairwise total-variation (TVpair) collapse indicators. Under stratified cross-validation, SPATE yields stronger representations across multiple datasets. For example, SPATE reaches a CKTA of 0.966 and a Fisher score of 7.37 on Blobs, compared with a CKTA of 0.632 and a Fisher score of 0.70 using angle encoding, and achieves a CKTA of 0.506 on Moons, compared with 0.015 using angle or amplitude encoding. These gains translate into stronger hybrid quantum neural network performance within a fixed qubit budget across several tasks, including an accuracy of 0.826 and an AUC of 0.978 for Wine, as well as an accuracy of 0.840 and an AUC of 0.923 for Moons. These results demonstrate that SPATE provides a practical spike-to-phase interface for building more informative quantum feature representations under constrained resources.
\end{abstract}


\section{Introduction}
Quantum Machine Learning (QML) models operate on quantum states \cite{biamonte2017quantum,cerezo2022challenges}, which makes state preparation one of the most critical steps in the pipeline: it determines how a classical sample is embedded into Hilbert space and therefore what information becomes accessible to a model under a limited qubit and circuit budget. The prepared state is either measured to produce a probability-vector embedding for representation analysis or for input to a classical learner, or used as the input layer of a trainable QML model, such as a Quantum Neural Network (QNN) \cite{schuld2014quest,innan2025next,innan2025qnn,innan2025fl,innan2025qfnn,innan2025quantum,innan2025lep}. In both settings, the encoding stage fixes the initial representation geometry on which learning and generalization depend; if samples from different classes are not well organized after encoding, subsequent training has limited capacity to recover this structure with shallow circuits and few qubits.

Standard encodings such as angle and amplitude are widely used because they are simple and broadly supported \cite{schuld2019quantum,rath2024quantum,dave2025sentiqnf, alami2025fid,ahmad2026quantum}, but their constraints become pronounced in realistic, capacity-limited regimes. Angle encoding maps features to local rotations, which can be expressive yet may not yield globally class-consistent structure when the number of qubits is small. Amplitude encoding represents a real vector through $2^n$ amplitudes on $n$ qubits, requiring a length-$2^n$ input and $\ell_2$ normalization; measurement probabilities then reflect squared normalized magnitudes. These properties can produce embeddings with weak label alignment or limited separability, which can directly limit the effectiveness of task-driven QNN training.

We address this limitation by introducing SPATE (Spiking-Phase Adaptive Temporal Encoding), a spike-informed state preparation that constructs circuit parameters from spike-derived statistics. SPATE converts each bounded input into per-feature spike rate and spike-timing phase using a lightweight leaky integrate-and-fire (LIF) mechanism, and it injects coarse temporal patterns by coupling feature qubits to a compact register of time qubits through controlled phase rotations. This design encodes intensity, timing, and coarse temporal structure within a fixed qubit budget and shallow circuits.

\begin{figure}[htpb]
    \centering
    \includegraphics[width=1\linewidth]{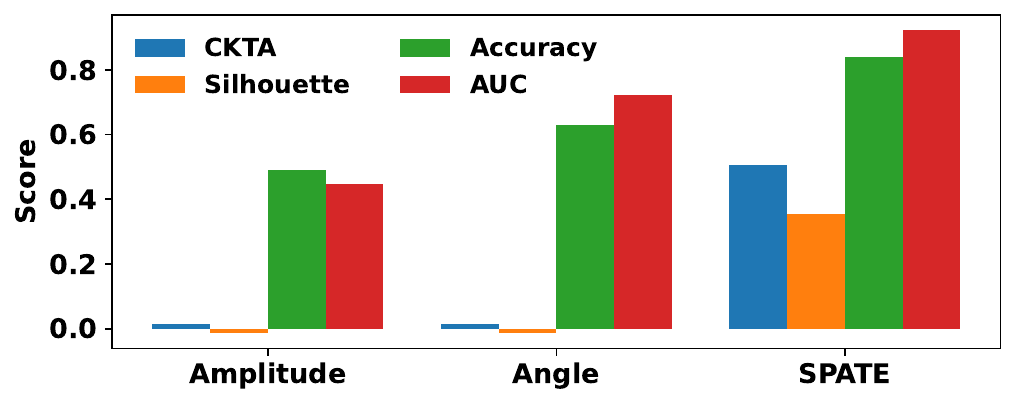}
\caption{Motivation example on Moons: SPATE yields a more class-consistent embedding geometry than standard angle and amplitude encodings, and this shift is reflected in hybrid-QNN classification performance under the same training setup.}
    \label{fig:motivation_moons}
\end{figure}

To motivate SPATE, we analyze measurement-based embedding geometry \emph{before} adding any trainable QNN layers. On a representative dataset (Moons), SPATE produces a more class-consistent embedding than standard angle and amplitude encodings under the same protocol, and this difference is reflected in classification performance when the same hybrid QNN classifier is applied (see Fig.~\ref{fig:motivation_moons}). This example supports the paper's premise: improving the organization of the encoded state can make limited-capacity QNNs easier to train and more effective, as learning starts from an embedding that better reflects the class structure.

Our contributions are:
\begin{itemize}
    \item We introduce \textbf{SPATE}, a spike-informed state-preparation method that encodes spike rate, spike-timing phase, and coarse temporal signatures using shallow rotations on feature qubits and controlled phase couplings to time qubits.
    \item We provide an \textbf{encoder-level evaluation} against angle and amplitude encodings using representation diagnostics and embedding-geometry visualizations under matched protocols and comparable qubit budgets.
    \item We demonstrate \textbf{task-level impact on classification} in a fixed trainable hybrid-QNN classifier, and report regimes where standard encodings remain better matched to the data geometry.
\end{itemize}

\section{Preliminaries}

\subsection{Neural Encoding in Spiking Neural Networks (SNNs)}

Spiking Neural Networks (SNNs) are neural network algorithms that are inspired by how the brain works, i.e., employing spiking neurons and spike-based operations~\cite{Ref_Putra_QSpiNN_IJCNN21}.
To support SNN processing, proper data-to-spike conversion is required so that the spike sequences effectively capture both temporal and spatial information from the data. 
Toward this, several neural/spike encoding schemes have been proposed, such as rate- and temporal-based coding~\cite{Ref_Putra_FSpiNN_TCAD20}.
\begin{itemize}
    \item \textit{Rate coding} employs the frequency of spikes (spike counts) to encode data. For instance, a higher intensity data sample is converted into a higher spike count.
    \item \textit{Temporal coding} employs temporal information to encode data. It has several variants, such as rank-order and phase coding.
    \textit{Rank-order coding} encodes data based on the order of spikes, while \textit{phase coding} encodes data based on a reference oscillator to determine the timing of spikes. 
\end{itemize}

In this work, we show the potential of spike encoding to provide both temporal and spatial information for enhancing quantum computing performance.

\subsection{Related Work}
Data encoding in QML covers a broad range of strategies, ranging from standard baselines such as angle and amplitude encodings to application-driven designs that are either formulated as an explicit input representation stage or embedded within the model through trainable feature maps and data re-uploading. Since the encoding fixes how classical structure is embedded into Hilbert space, it effectively determines the initial representation geometry available to learning under realistic constraints (few qubits, shallow circuits, limited measurement budgets). This motivates a growing body of work at the interface of neuromorphic computing and quantum learning, where spiking representations are used to inject temporal structure that is difficult to capture with static encoders.

A first line of work uses SNNs as a feature extraction stage to derive spike-based features from spatio-temporal signals, followed by a quantum kernel or quantum classifier that operates on these features \cite{jha2025hybrid,jha2025spiking}. These frameworks demonstrate that spike timing or spike-rate summaries can preserve temporal interactions and yield improved classification relative to hand-crafted statistics, but they typically treat the spiking stage as a separate learned feature extractor and do not isolate the encoding circuit itself as the object of study. Another direction introduces quantum-native spiking primitives and architectures, including quantum LIF neurons and quantum spiking neural networks (QSNNs), as well as measurement strategies intended to reduce shot overhead and make spiking computation more compatible with near-term devices \cite{brand2024quantum,liu2025quantum}. More recent proposals highlight remaining limitations of existing quantum spiking models, such as reliance on repeated measurements for firing estimation or training procedures anchored to classical backpropagation, and propose stochastic quantum spiking units with internal quantum memory and more hardware-aligned learning rules \cite{chen2025stochastic}. In addition, end-to-end hybrid designs that combine spiking components with quantum layers via data re-uploading aim to improve differentiability and joint optimization, but they often require specialized architectural coupling and introduce training and scalability constraints that make it difficult to attribute improvements specifically to the encoding mechanism \cite{nhan2025parameter}.

Despite this progress, a gap remains between neuromorphic-quantum integration and the specific question of state preparation as a modular encoding stage for general QML pipelines: existing SNN-quantum frameworks emphasize learned spiking feature extraction, while QSNN-style approaches emphasize neuron/network construction and training methodology. SPATE targets this gap by treating spiking dynamics as a lightweight parameter generator rather than a standalone trainable spiking model. It converts bounded inputs into spike-derived rate and spike-timing phase, and encodes coarse temporal structure through controlled phase interactions in a shallow circuit, enabling the same encoding block to be attached to different QML models and tasks without committing to a dedicated QSNN architecture or an SNN pretraining stage.

\section{Methodology} \label{method}

\begin{figure*}[htpb]
    \centering
    \includegraphics[width=0.95\linewidth]{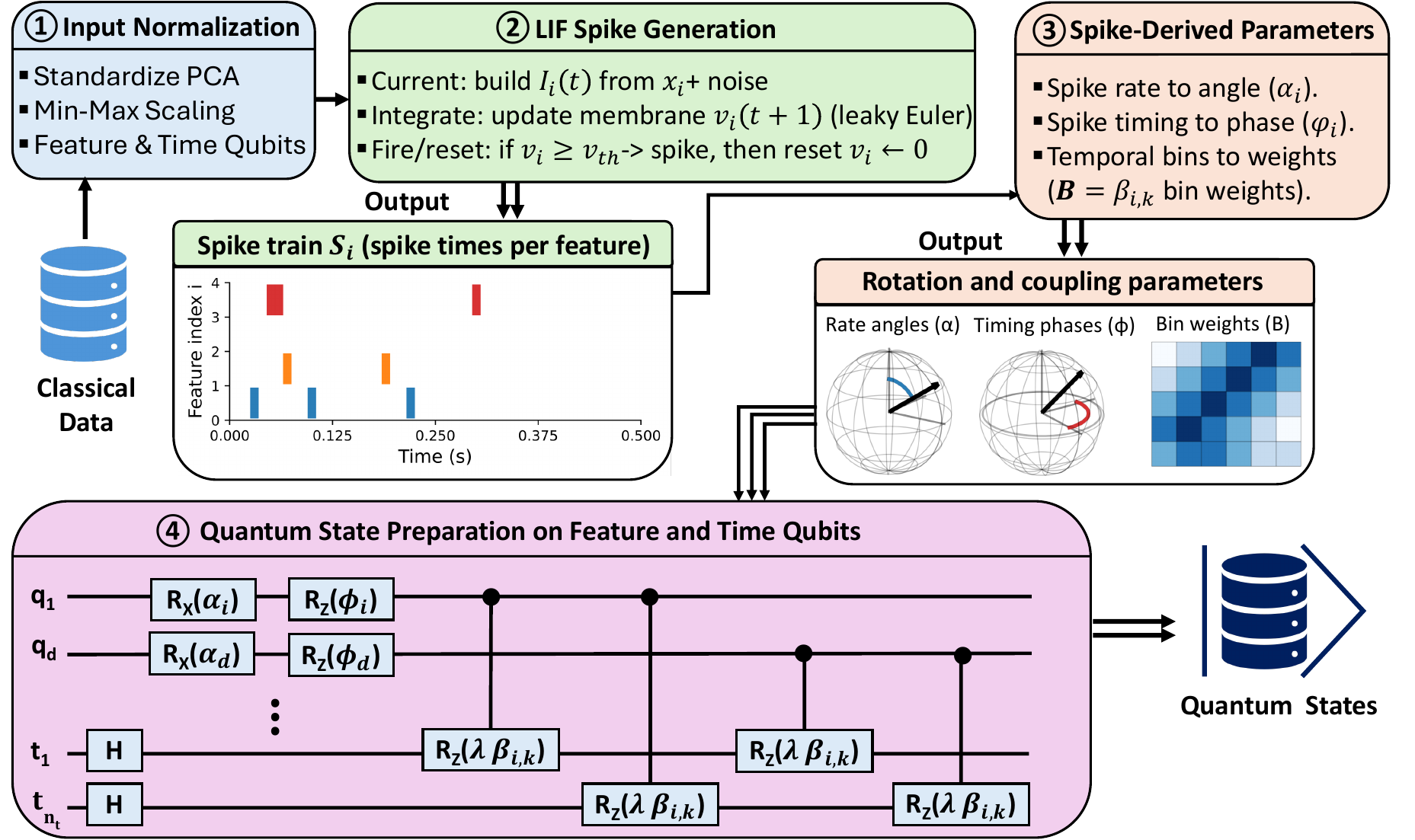}
    \vspace{0.2cm}
    \caption{SPATE methodology pipeline. From a standardized and $[0,1]$-scaled input, SPATE generates per-feature spike trains (LIF), extracts spike rate $\boldsymbol{\alpha}$, timing phase $\boldsymbol{\phi}$, and temporal-bin signatures $\mathbf{B}$, then prepares the quantum state using feature rotations and feature-to-time $\mathrm{CRZ}(\lambda\,\beta_{i,k})$ couplings.}
    \label{figmeth}
\end{figure*}

To construct a quantum state that carries feature intensity information, spike-timing phase information, and coarse temporal patterns, SPATE maps each input amplitude to a set of rotation angles derived from a LIF spike generator, then prepares a quantum state on feature and time qubits using single-qubit rotations and controlled phase interactions as shown in Fig. \ref{figmeth}.

\subsection{Input Normalization}
To ensure a bounded and consistent spike-generation regime, each amplitude is first standardized and then scaled feature-wise to $[0,1]$. Let $\mathbf{x}\in\mathbb{R}^{d}$ denote the standardized feature vector after PCA (so $d=d_{\text{enc}}$), and let $\tilde{\mathbf{x}}\in[0,1]^d$ be the MinMax-scaled version used by SPATE. SPATE uses $d$ feature qubits and $n_t$ time qubits, for a total of $n_{\text{SPATE}}=d+n_t$, and maps each input to a probability-vector embedding $\mathbf{p}\in\Delta^{2^{n_{\text{SPATE}}}}$ obtained by measuring the prepared state in the computational basis.
\subsection{LIF Spike Generation (per feature)}
To convert each normalized feature $\tilde{x}_i$ into a spike train, SPATE simulates a discrete-time LIF neuron over a horizon $T$ with step size $\Delta t$, for $N=\lfloor T/\Delta t\rfloor$ updates. For feature $i\in\{1,\dots,d\}$, the membrane potential $v_i$ is updated as
\begin{equation}
v_i(t+\Delta t) = v_i(t) + \frac{\Delta t}{\tau}\Big(-v_i(t) + g\,\tilde{x}_i + \sigma\,\epsilon_i(t)\Big),
\label{eq:lif_update}
\end{equation}
where $\tau>0$ is the membrane time constant, $g>0$ is a gain factor, $\sigma\ge 0$ controls additive noise, and $\epsilon_i(t)\sim\mathcal{N}(0,1)$ is i.i.d. Gaussian noise (seeded for reproducibility). If $v_i(t+\Delta t)\ge v_{\text{th}}$, SPATE records a spike at time $t+\Delta t$ and resets $v_i(t+\Delta t)\leftarrow 0$. Let $\mathcal{S}_i=\{s_{i,1},\dots,s_{i,c_i}\}$ be the set of spike times for feature $i$ within $[0,T]$, and let $c_i = |\mathcal{S}_i|$ be the spike count.
\subsection{Spike-Derived Parameters (rate, phase, temporal bins)}
To convert spike trains into circuit parameters, SPATE extracts three components from $\mathcal{S}_i$.

To encode spike rate as an amplitude-like rotation, SPATE maps the spike count to an angle $\alpha_i\in[0,\pi]$:
\begin{equation}
N=\left\lfloor \frac{T}{\Delta t}\right\rfloor,\qquad
c_{\max}=N,\qquad
\alpha_i = \pi\sqrt{\frac{c_i}{c_{\max}+\varepsilon}},
\label{eq:alpha_rate}
\end{equation}
where $c_{\max}$ is a conservative upper bound on the number of spikes over the simulation horizon and $\varepsilon$ is a small constant for numerical stability.

To encode spike timing as a circular phase, SPATE computes the circular mean of spike phases on $[0,T]$. Define the phase of a spike time $s$ as $\theta(s)=2\pi s/T$. The feature phase $\phi_i\in[0,2\pi)$ is
\begin{equation}
\phi_i =
\begin{cases}
\arg\Big(\frac{1}{c_i}\sum\limits_{m=1}^{c_i} e^{j\,\theta(s_{i,m})}\Big)\ \bmod\ 2\pi, & c_i>0,\\
0, & c_i=0.
\end{cases}
\label{eq:phi_phase}
\end{equation}

To encode coarse temporal patterns, SPATE discretizes $[0,T]$ into $n_t$ bins with edges $\{0,\frac{T}{n_t},\dots,T\}$. Let $h_{i,k}$ be the histogram count of spikes of feature $i$ in bin $k\in\{1,\dots,n_t\}$. SPATE then centers the histogram per feature to remove global offsets $\beta_{i,k} = h_{i,k} - \frac{1}{n_t}\sum\limits_{k'=1}^{n_t} h_{i,k'}.$

Note that $\beta_{i,k}$ can be signed; signed values are retained and directly used as phase-rotation parameters in the temporal-coupling step. The triple $(\boldsymbol{\alpha},\boldsymbol{\phi},\mathbf{B})$ with $\boldsymbol{\alpha}=(\alpha_1,\dots,\alpha_d)$, $\boldsymbol{\phi}=(\phi_1,\dots,\phi_d)$, and $\mathbf{B}=[\beta_{i,k}]_{i,k}$ fully defines SPATE's state-preparation angles for the sample.
\subsection{Quantum State Preparation on Feature and Time Qubits}
To inject temporal structure explicitly, SPATE uses $d$ feature qubits $\{q_1,\dots,q_d\}$ and $n_t$ time qubits $\{t_1,\dots,t_{n_t}\}$. Time qubits act as phase receptors for coarse temporal-bin signatures, rather than as a classical index register. For an input $\tilde{\mathbf{x}}$, the state is prepared as follows.

\noindent\textbf{Temporal reference.}
SPATE initializes each time qubit in the $|+\rangle$ state $|0\rangle^{\otimes n_t}\ \xrightarrow{\ H^{\otimes n_t}\ }\ |+\rangle^{\otimes n_t}.$

\noindent\textbf{Feature rate and phase.}
For each feature qubit $q_i$, SPATE encodes spike-derived rate and timing phase via $\forall i\in\{1,\dots,d\}:$ 
$R_X(\alpha_i)\ \text{on } q_i,$ and $R_Z(\phi_i)\ \text{on } q_i.$

\noindent\textbf{Feature-to-time coupling.}
To couple feature activity to temporal bins, for each $i\in\{1,\dots,d\}$ and $k\in\{1,\dots,n_t\}$, SPATE applies a controlled phase rotation $\mathrm{CRZ}\big(\lambda\,\beta_{i,k}\big)$ with control $q_i$ and target $t_k$, where $\lambda>0$ is a fixed scaling factor (beta\_scale). This step injects a signed, bin-dependent phase imprint conditioned on feature-level spiking patterns.

\noindent\textbf{Measurement and downstream usage.}
SPATE can be used in two modes. \emph{Embedding mode:} the prepared state $|\psi(\tilde{\mathbf{x}})\rangle$ is measured in the computational basis to obtain a probability-vector embedding $\mathbf{p}(\tilde{\mathbf{x}}) = \big(p_z\big)_{z\in\{0,1\}^{d+n_t}},$ and 
$p_z = \left|\langle z|\psi(\tilde{\mathbf{x}})\rangle\right|^2.$
This embedding is used for representation-level metrics and can also be provided as a fixed-dimensional input to downstream classical or hybrid models.  \emph{End-to-end QML mode:} SPATE serves as a data-encoding prefix circuit, after which a trainable QML ansatz is appended and optimized using task-driven readout measurements; in this mode, full probability-vector extraction is not required and is used only for encoder-level evaluation.

\subsection{Seed-Averaging (stochastic stability)}
To reduce variance introduced by the noise term in the LIF dynamics, SPATE repeats the spike-to-parameter extraction with multiple random seeds and averages the resulting measurement distributions. For $S$ seeds, the final embedding is the mean probability vector $\bar{\mathbf{p}}(\tilde{\mathbf{x}}) = \frac{1}{S}\sum_{s=1}^{S}\mathbf{p}^{(s)}(\tilde{\mathbf{x}}),$
which can be interpreted as a mixture embedding across seed-conditioned circuits.

\subsection{Fold-Wise Tuning (train-only)}
To avoid selecting SPATE parameters on test data, any tuning is performed using only the training split of each cross-validation fold. Let $\Theta$ be a small parameter grid over $(g,\sigma,v_{\text{th}},\tau)$ (and any fixed normalization constants, if applicable). For each candidate $\theta\in\Theta$, a score is computed on the training subset using the same representation objectives reported in evaluation (e.g., alignment and separability), and the best parameter set is selected: $\theta^\star = \arg\max_{\theta\in\Theta}\ \mathcal{J}(\theta;\ \text{train fold}).$
The selected $\theta^\star$ is then used to encode the corresponding test split of that fold. For fair comparison, any tunable hyperparameters of baseline encoders are selected under the same train-only protocol and with comparable grid sizes.

\subsection{Evaluation Protocol and Metrics}
\label{subsec:evaluation}
To quantify representation quality in a reproducible and model-agnostic way, each encoder maps an input sample to a probability-vector embedding $\mathbf{p}_i\in\mathbb{R}^{D}$ obtained from computational-basis measurement, with $D=2^{n}$ and $n$ the number of qubits used by the encoder. Unless stated otherwise, encoders are compared under a matched qubit budget (same $n$) to keep embedding dimension $D$ fixed across methods; when $n$ differs, we report $n$ explicitly and interpret metrics that depend on $D$ accordingly. All representation metrics are computed directly in this embedding space and aggregated under stratified $K$-fold cross-validation.

\subsubsection{Label Alignment}
To measure whether similarity in embedding space matches label similarity, we compute centered kernel target alignment (CKTA) between an RBF kernel built on $\{\mathbf{p}_i\}_{i=1}^N$ and a label kernel \cite{cortes2012algorithms}. Define $K_{ij}=\exp\!\big(-\gamma\|\mathbf{p}_i-\mathbf{p}_j\|_2^2\big),$ and 
$Y_{ij}=\mathbb{I}[y_i=y_j],$
where $\gamma$ is set using a median-distance heuristic. With the centering matrix $H=I-\frac{1}{N}\mathbf{1}\mathbf{1}^\top$, let $K_c=HKH$ and $Y_c=HYH$. CKTA is $\mathrm{CKTA}(K,Y)=\frac{\langle K_c,Y_c\rangle_F}{\|K_c\|_F\,\|Y_c\|_F+\varepsilon},$
with $\langle A,B\rangle_F=\sum_{ij}A_{ij}B_{ij}$.

\subsubsection{Separability (Fisher, Inter/Intra, Silhouette)}
To measure class separation in the embedding space, we report Fisher ratio, inter-/intra-class distance ratio, and the Silhouette score computed on $\{\mathbf{p}_i\}$ \cite{duda1973pattern,rousseeuw1987silhouettes}.

Let $\boldsymbol{\mu}$ be the global mean of embeddings, $\boldsymbol{\mu}_c$ the mean of class $c$, and $n_c$ the number of samples in class $c$. The between-class and within-class scatters are $S_B=\sum_{c} n_c\|\boldsymbol{\mu}_c-\boldsymbol{\mu}\|_2^2,$ and
$S_W=\sum_{c}\sum_{i:y_i=c}\|\mathbf{p}_i-\boldsymbol{\mu}_c\|_2^2,$
and the Fisher ratio is $\mathrm{Fisher}=\frac{S_B}{S_W+\varepsilon}.$

For the inter-/intra-class ratio, let $D_{ij}=\|\mathbf{p}_i-\mathbf{p}_j\|_2$. Define $\overline{D}_{\text{intra}}=\mathbb{E}[D_{ij}\mid y_i=y_j],\qquad
\overline{D}_{\text{inter}}=\mathbb{E}[D_{ij}\mid y_i\ne y_j],$
(excluding diagonal terms), and report $\mathrm{Inter/Intra}=\frac{\overline{D}_{\text{inter}}}{\overline{D}_{\text{intra}}+\varepsilon}.$

For Silhouette, for each point $i$ let $a(i)$ be the average distance from $i$ to points in its own class and let $b(i)$ be the minimum average distance from $i$ to points in any other class. Then
\begin{equation}
s(i)=\frac{b(i)-a(i)}{\max\{a(i),b(i)\}+\varepsilon},\qquad
\mathrm{Sil}=\frac{1}{N}\sum_{i=1}^{N}s(i).
\label{eq:silhouette}
\end{equation}

\subsubsection{Information and Non-Collapse (Hnorm, TVpair)}
To quantify distribution richness and guard against representational collapse, we report normalized Shannon entropy and mean pairwise total variation distance \cite{shannon1948mathematical,ijcai2023p387}.

Given $\mathbf{p}\in\mathbb{R}^{D}$, Shannon entropy (base 2) is
\begin{equation}
H(\mathbf{p})=-\sum_{z=1}^{D} p_z\log_2(p_z+\varepsilon),
\label{eq:entropy}
\end{equation}
and we normalize by the number of qubits $n$ to compare encodings with different Hilbert-space sizes $H_{\text{norm}}=\frac{H(\mathbf{p})}{n}.$

For non-collapse, we compute the total variation distance between two embeddings $\mathrm{TV}(\mathbf{p},\mathbf{q})=\frac{1}{2}\|\mathbf{p}-\mathbf{q}\|_1
=\frac{1}{2}\sum_{z=1}^{D}|p_z-q_z|,$
and report TVpair as the mean TV over a fixed number of randomly sampled pairs $(i,j)$ from the evaluation split $\mathrm{TVpair}=\mathbb{E}_{(i,j)}\big[\mathrm{TV}(\mathbf{p}_i,\mathbf{p}_j)\big].$
When $n$ differs across encoders, we additionally report the normalized variant $\mathrm{TVpair}_{\text{norm}}=\mathrm{TVpair}/n$ to reduce dimension-induced effects.

\subsubsection{Downstream hybrid-QNN Evaluation}
To test whether improved embeddings yield better learning behavior, we train a hybrid QNN classifier with a fixed architecture and optimization settings across encoders. The model outputs a readout probability vector $\mathbf{r}_i\in\mathbb{R}^{2^m}$ from $m=\lceil\log_2(C)\rceil$ readout qubits, then maps it to $C$ class probabilities by truncation and renormalization:
\begin{equation}
\hat{p}_{i,c}=\frac{r_{i,c}}{\sum_{c'=1}^{C} r_{i,c'}+\varepsilon},\qquad c\in\{1,\dots,C\}.
\label{eq:readout_map}
\end{equation}
Training minimizes cross-entropy over the training split:
\begin{equation}
\mathcal{L} = -\frac{1}{N}\sum_{i=1}^{N}\sum_{c=1}^{C}\mathbb{I}[y_i=c]\log(\hat{p}_{i,c}+\varepsilon).
\label{eq:cross_entropy}
\end{equation}
We report accuracy, precision, recall (macro for $C>2$ and binary for $C=2$), and AUC (binary AUC for $C=2$, macro one-vs-rest AUC for $C>2$), averaged across folds.
\begin{table*}[b]
\centering
\caption{Encoding quality metrics averaged over 5-fold stratified cross-validation (mean$\pm$std over folds). $H_{\text{norm}}$ is Shannon entropy normalized by the number of qubits ($H/n_{\text{qubits}}$), enabling fair comparison across encodings with different Hilbert-space sizes. TVpair is the mean pairwise total variation distance (non-collapse indicator).}
\small
\begin{tabular}{llcccccc}
\hline
Dataset & Enc & CKTA $\uparrow$ & Fisher $\uparrow$ & Inter/Intra $\uparrow$ & Sil $\uparrow$ & $H_{\text{norm}}$ $\uparrow$ & TVpair $\uparrow$\\
\hline
Iris   & Amplitude  & 0.2910$\pm$0.0848 & 0.3381$\pm$0.1603 & 1.2144$\pm$0.1133 & -0.0078$\pm$0.0500 & 0.3727$\pm$0.0116 & 0.4406$\pm$0.0189\\
Iris   & Angle      & 0.5034$\pm$0.0358 & 0.7458$\pm$0.1013 & 1.4870$\pm$0.0760 & 0.1414$\pm$0.0390 & 0.5606$\pm$0.0296 & 0.4973$\pm$0.0217 \\
Iris   & SPATE      & \cellcolor{green!18}0.7436$\pm$0.0127 & \cellcolor{green!18}3.2175$\pm$0.8327 & \cellcolor{green!18}2.4596$\pm$0.2454 & \cellcolor{green!18}0.3646$\pm$0.0333 & \cellcolor{green!18}0.7828$\pm$0.0137 & \cellcolor{green!18}0.6359$\pm$0.0149  \\
\hline
Wine   & Amplitude  & 0.4132$\pm$0.0585 & 0.4947$\pm$0.1414 & 1.2589$\pm$0.0940 & 0.0245$\pm$0.0344 & 0.2211$\pm$0.0038 & 0.5215$\pm$0.0247 \\
Wine   & Angle      & 0.3810$\pm$0.0544 & 0.2975$\pm$0.0617 & 1.1617$\pm$0.0509 & 0.0555$\pm$0.0378 & \cellcolor{green!18}0.4529$\pm$0.0212 & \cellcolor{green!18}0.7442$\pm$0.0129 \\
Wine   & SPATE      & \cellcolor{green!18}0.6597$\pm$0.0200 & \cellcolor{green!18}0.8122$\pm$0.0571 & \cellcolor{green!18}1.5324$\pm$0.0447 & \cellcolor{green!18}0.2925$\pm$0.0191 & 0.4186$\pm$0.0092 & 0.5035$\pm$0.0185  \\
\hline
Cancer & Amplitude  & 0.0480$\pm$0.0103 & 0.0223$\pm$0.0069 & 1.0115$\pm$0.0183 & 0.0128$\pm$0.0125 & 0.2128$\pm$0.0030 & 0.5062$\pm$0.0071\\
Cancer & Angle      & 0.0969$\pm$0.0224 & 0.0337$\pm$0.0095 & 1.0242$\pm$0.0087 & 0.0212$\pm$0.0038 & 0.4756$\pm$0.0126 & \cellcolor{green!18}0.8163$\pm$0.0040\\
Cancer & SPATE      & \cellcolor{green!18}0.2718$\pm$0.0264 & \cellcolor{green!18}0.0832$\pm$0.0107 & \cellcolor{green!18}1.1156$\pm$0.0245 & \cellcolor{green!18}0.0966$\pm$0.0145 & \cellcolor{green!18}0.8151$\pm$0.0069 & 0.5993$\pm$0.0194 \\
\hline
Digits & Amplitude  & 0.3027$\pm$0.0124 & 0.6988$\pm$0.0699 & 1.3760$\pm$0.0337 & -0.0334$\pm$0.0121 & 0.2402$\pm$0.0032 & 0.5969$\pm$0.0063\\
Digits & Angle      & 0.2126$\pm$0.0138 & 0.1265$\pm$0.0109 & 1.0621$\pm$0.0062 & -0.0194$\pm$0.0124 & 0.5456$\pm$0.0041 & \cellcolor{green!18}0.8873$\pm$0.0038\\
Digits & SPATE      & \cellcolor{green!18}0.4922$\pm$0.0249 & \cellcolor{green!18}0.7941$\pm$0.0591 & \cellcolor{green!18}1.4232$\pm$0.0268 & \cellcolor{green!18}0.1019$\pm$0.0184 & \cellcolor{green!18}0.6917$\pm$0.0324 & 0.6964$\pm$0.0076\\
\hline
Moons  & Amplitude  & 0.0145$\pm$0.0083 & 0.0061$\pm$0.0024 & 0.9839$\pm$0.0061 & -0.0132$\pm$0.0074 & 0.2573$\pm$0.0166 & 0.4203$\pm$0.0076\\
Moons  & Angle      & 0.0145$\pm$0.0060 & 0.0038$\pm$0.0016 & 0.9844$\pm$0.0037 & -0.0141$\pm$0.0044 & 0.5658$\pm$0.0160 & 0.3283$\pm$0.0082\\
Moons  & SPATE      & \cellcolor{green!18}0.5057$\pm$0.0874 & \cellcolor{green!18}0.7814$\pm$0.1734 & \cellcolor{green!18}1.6977$\pm$0.1461 & \cellcolor{green!18}0.3554$\pm$0.0569 & \cellcolor{green!18}0.8574$\pm$0.0073 & \cellcolor{green!18}0.4868$\pm$0.0124 \\
\hline
Circles& Amplitude  & 0.0293$\pm$0.0338 & 0.0287$\pm$0.0374 & 1.0116$\pm$0.0479 & 0.0086$\pm$0.0388 & 0.2761$\pm$0.0231 & \cellcolor{green!18}0.4030$\pm$0.0219\\
Circles& Angle      & \cellcolor{green!18}0.8176$\pm$0.0262 & \cellcolor{green!18}1.0698$\pm$0.1219 & \cellcolor{green!18}2.2572$\pm$0.1075 & \cellcolor{green!18}0.5539$\pm$0.0220 & 0.5687$\pm$0.0100 & 0.3287$\pm$0.0077 \\
Circles& SPATE      & 0.2284$\pm$0.0513 & 0.1976$\pm$0.0632 & 1.2435$\pm$0.0824 & 0.1732$\pm$0.0476 & \cellcolor{green!18}0.6785$\pm$0.0063 & 0.1542$\pm$0.0080\\
\hline
Blobs  & Amplitude  & 0.7192$\pm$0.0210 & 1.3296$\pm$0.0891 & 1.7719$\pm$0.0437 & 0.3514$\pm$0.0183 & 0.3595$\pm$0.0091 & 0.4675$\pm$0.0183\\
Blobs  & Angle      & 0.6322$\pm$0.0174 & 0.7045$\pm$0.0348 & 1.4558$\pm$0.0141 & 0.2541$\pm$0.0134 & 0.6087$\pm$0.0153 & 0.4923$\pm$0.0113\\
Blobs  & SPATE      & \cellcolor{green!18}0.9661$\pm$0.0050 & \cellcolor{green!18}7.3651$\pm$0.9566 & \cellcolor{green!18}3.8042$\pm$0.2300 & \cellcolor{green!18}0.7070$\pm$0.0189 & \cellcolor{green!18}0.7009$\pm$0.0067 & \cellcolor{green!18}0.6397$\pm$0.0066\\
\hline
\end{tabular}
\label{tab:encoding_metrics_cv}
\end{table*}
\section{Results and Discussion}
\label{subsec:results_discussion}
\subsection{Experimental Setup}
We compare SPATE against two standard baselines: angle encoding and amplitude encoding. All results use stratified 5-fold cross-validation with a fixed global seed and statevector simulation on \texttt{default.qubit}, so outputs correspond to exact computational-basis probabilities.

Datasets are standardized (zero mean, unit variance). When the standardized feature dimension exceeds 8, PCA is applied to obtain $d_{\text{enc}}=\min(d,8)$ (Table~\ref{tab:datasets_summary}). In the downstream hybrid-QNN study, PCA is additionally used as needed to match the available feature-qubit capacity of each encoder.
\begin{table}[htpb]
    \centering
\caption{Datasets and preprocessing used in the encoding-quality study. $d_{\text{enc}}$ is the effective feature dimension after standardization and PCA capping to 8. Synthetic datasets are generated with a fixed random seed.}
\label{tab:datasets_summary}
\small
\begin{tabular}{lcccc}
\hline
Dataset & \#Samples & $d$ (orig.) & \#Classes & $d_{\text{enc}}$ \\
\hline
Iris    & 150  & 4  & 3  & 4 \\
Wine    & 178  & 13 & 3  & 8 \\
Cancer  & 569  & 30 & 2  & 8 \\
Digits  & 1797 & 64 & 10 & 8 \\
Moons   & 300  & 2  & 2  & 2 \\
Circles & 300  & 2  & 2  & 2 \\
Blobs   & 300  & 5  & 3  & 5 \\
\hline
\end{tabular}
\end{table}
For the baselines, angle encoding applies \texttt{AngleEmbedding} on $n$ encoder qubits and returns a $2^n$-dimensional measurement distribution. Amplitude encoding prepares an $n$-qubit amplitude-encoded state; since this requires an input vector of length $2^n$, each standardized sample is padded with zeros or truncated to length $2^n$ and then $\ell_2$-normalized before state preparation.

For the downstream hybrid-QNN study, we fix the architecture and optimization settings across encoders: \texttt{StronglyEntanglingLayers} with 2 layers, Adam, 120 training steps, batch size 32, learning rate 0.15, and cross-entropy loss. The readout uses $m=\lceil\log_2(C)\rceil$ qubits; class probabilities are obtained by truncating the $2^m$ readout distribution to the first $C$ entries and renormalizing. To keep the total qubit budget comparable, angle/amplitude encodings use up to 6 feature qubits, while SPATE uses $n_t{=}3$ time qubits plus up to 3 feature qubits (minimum 2).

\subsection{Encoding Quality and Similarity Analysis}
\label{subsubsec:enc_quality}

To assess whether the proposed SPATE state preparation produces representations that are more class-aligned and more separable than angle and amplitude under the same qubit budget, Table~\ref{tab:encoding_metrics_cv} reports similarity, separability, and distribution diagnostics across datasets. SPATE achieves higher alignment and separability on most benchmarks. On \textbf{Iris}, CKTA increases to $0.7436\pm0.0127$ compared to $0.5034\pm0.0358$ (angle) and $0.2910\pm0.0848$ (amplitude), while Fisher increases to $3.2175\pm0.8327$ compared to $0.7458\pm0.1013$ (angle), indicating a stronger class-consistent structure. On \textbf{Moons}, SPATE yields a marked shift from weakly structured embeddings to clearly clusterable ones: CKTA rises from $\approx 0.0145$ (angle/amplitude) to $0.5057\pm0.0874$ and Silhouette changes from negative values to $0.3554\pm0.0569$. On \textbf{Blobs}, SPATE achieves the strongest separability across metrics (CKTA $0.9661\pm0.0050$, Fisher $7.3651\pm0.9566$, Silhouette $0.7070\pm0.0189$), consistent with clean inter-class margins.

To verify that these gains are not driven by representation collapse and to compare encodings with different Hilbert-space sizes, Table~\ref{tab:encoding_metrics_cv} additionally reports $H_{\text{norm}}$ and TVpair. SPATE tends to increase $H_{\text{norm}}$, indicating richer state distributions per qubit. On \textbf{Cancer}, $H_{\text{norm}}$ increases to $0.8151\pm0.0069$ compared to $0.4756\pm0.0126$ (angle) and $0.2128\pm0.0030$ (amplitude), while TVpair remains non-trivial at $0.5993\pm0.0194$, suggesting that SPATE expands representation richness while preserving meaningful sample-to-sample variability. The main exception is \textbf{Circles}, where angle best matches the underlying geometry (CKTA $0.8176\pm0.0262$, Silhouette $0.5539\pm0.0220$), while SPATE is lower (CKTA $0.2284\pm0.0513$), indicating that simple rotation-based periodic mappings can be particularly well-suited to ring-shaped decision boundaries.

\subsection{t-SNE Visualization of Embedding Geometry}
\begin{figure*}[htpb]
    \centering
    \includegraphics[width=0.95\linewidth]{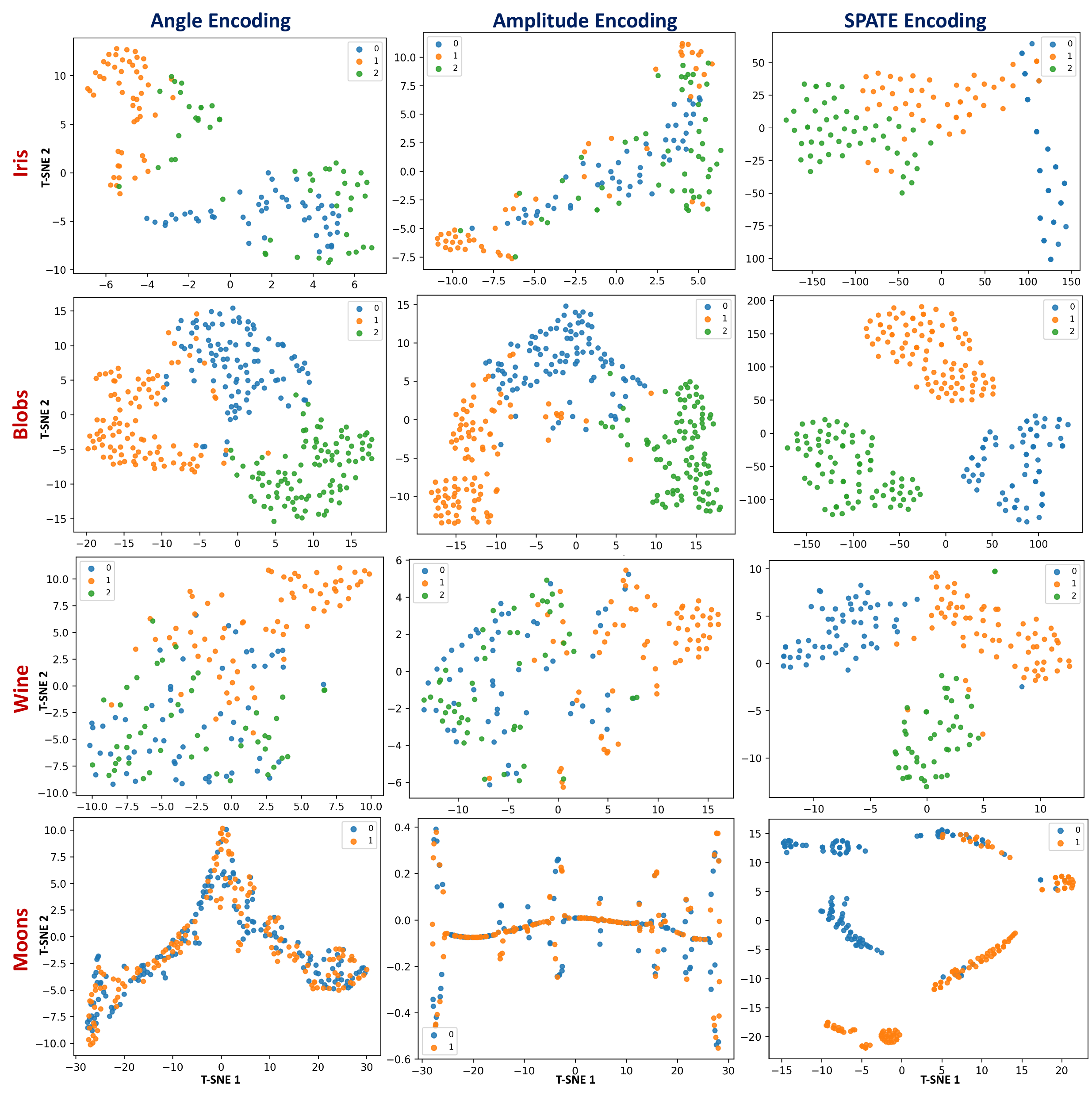}
    \caption{t-SNE projections of the embedding space for Iris, Wine, Moons, and Blobs datasets.}
    \label{fig:tsne_a}
\end{figure*}

\begin{figure*}[htpb]
    \centering
    \includegraphics[width=0.94\linewidth]{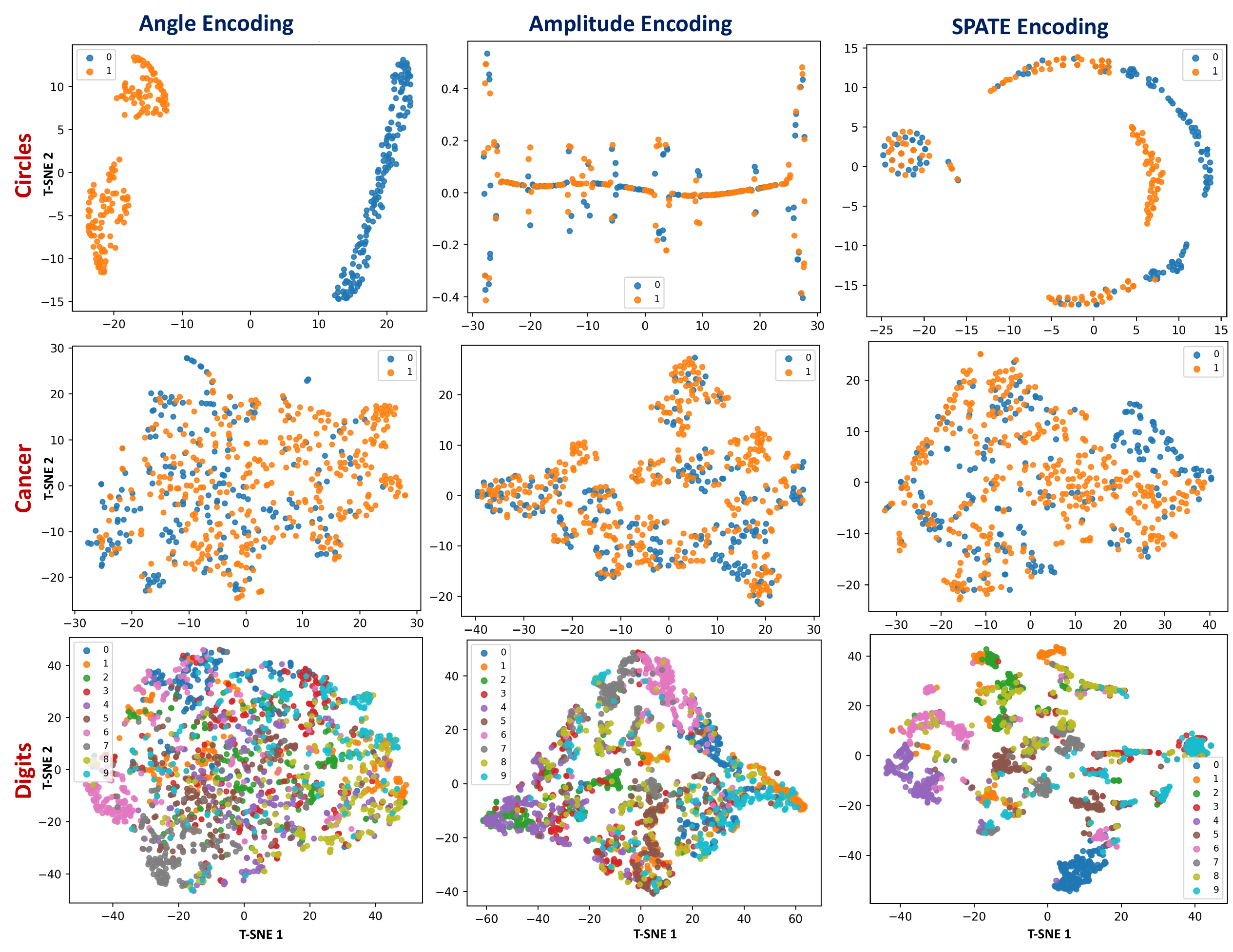}
    \caption{t-SNE projections of the embedding space for Cancer, Circles, and Digits datasets.}
    \label{fig:tsne_b}
\end{figure*}

To provide an intuitive view of how each encoding organizes samples before introducing the trainable QNN, we visualize the resulting embeddings using t-SNE (Figs.~\ref{fig:tsne_a}--\ref{fig:tsne_b}). In Fig.~\ref{fig:tsne_a}, SPATE produces the clearest neighborhood structure in the same datasets where Table~\ref{tab:encoding_metrics_cv} reports the largest separability gains. On \textbf{Moons}, SPATE forms two separated manifolds with limited mixing, whereas amplitude compresses samples into an almost one-dimensional trace, and angle shows more overlap along the curve. On \textbf{Blobs}, SPATE yields three compact, well-separated clusters, while angle and amplitude exhibit more boundary mixing. On \textbf{Wine} and \textbf{Iris}, SPATE produces more distinct groupings than amplitude and typically reduces overlap compared to angle, aligning with the higher CKTA and Fisher values reported in Table~\ref{tab:encoding_metrics_cv}.

To highlight regimes where the baselines remain competitive or where separation is intrinsically limited in 2D, Fig.~\ref{fig:tsne_b} groups Cancer, Circles, and Digits. On \textbf{Circles}, angle shows the cleanest separation pattern, matching its dominant CKTA and Silhouette values. On \textbf{Digits}, overlap remains substantial for all encodings in 2D, which is expected in a 10-class setting under a tight qubit/readout budget; nevertheless, SPATE tends to form more localized islands with more homogeneous neighborhoods than angle, indicating improved local organization even when global separation remains difficult. On \textbf{Cancer}, all encodings show partial mixing in 2D, but SPATE exhibits more structured sub-regions than amplitude, which is consistent with its higher $H_{\text{norm}}$ and improved separability scores.

\subsection{Hybrid-QNN Classification Performance Analysis}
\label{subsubsec:qnn_perf}

\begin{table}[htbp]
\centering
\caption{Performance summary (mean $\pm$ std) across datasets and encodings.}
\label{tab:qnn_results_summary}
\small
\setlength{\tabcolsep}{3pt}
\renewcommand{\arraystretch}{1.05}
\resizebox{\columnwidth}{!}{%
\begin{tabular}{llcccc}
\hline
Dataset & Encoding & Accuracy & Precision & Recall & AUC \\
\hline
Iris   & Angle      & \cellcolor{green!18}$0.760 \pm 0.039$ & \cellcolor{green!18}$0.768 \pm 0.046$ & \cellcolor{green!18}$0.760 \pm 0.039$ & \cellcolor{green!18}$0.911 \pm 0.017$ \\
Iris   & Amplitude  & $0.573 \pm 0.077$ & $0.620 \pm 0.085$ & $0.573 \pm 0.077$ & $0.748 \pm 0.072$ \\
Iris   & SPATE      & $0.653 \pm 0.091$ & $0.746 \pm 0.084$ & $0.653 \pm 0.091$ & $0.890 \pm 0.036$ \\
\hline
Wine   & Angle      & $0.399 \pm 0.063$ & $0.398 \pm 0.072$ & $0.398 \pm 0.067$ & $0.552 \pm 0.050$ \\
Wine   & Amplitude  & $0.685 \pm 0.023$ & $0.717 \pm 0.042$ & $0.698 \pm 0.032$ & $0.887 \pm 0.026$ \\
Wine   & SPATE      & \cellcolor{green!18}$0.826 \pm 0.071$ & \cellcolor{green!18}$0.810 \pm 0.141$ & \cellcolor{green!18}$0.810 \pm 0.098$ & \cellcolor{green!18}$0.978 \pm 0.018$ \\
\hline
Cancer & Angle      & $0.599 \pm 0.067$ & $0.704 \pm 0.060$ & $0.622 \pm 0.065$ & $0.643 \pm 0.075$ \\
Cancer & Amplitude  & $0.620 \pm 0.070$ & $0.728 \pm 0.056$ & $0.628 \pm 0.096$ & $0.676 \pm 0.070$ \\
Cancer & SPATE      & \cellcolor{green!18}$0.837 \pm 0.062$ & \cellcolor{green!18}$0.806 \pm 0.062$ & \cellcolor{green!18}$0.983 \pm 0.011$ & \cellcolor{green!18}$0.772 \pm 0.112$ \\
\hline
Digits & Angle      & $0.307 \pm 0.074$ & $0.304 \pm 0.079$ & $0.309 \pm 0.075$ & $0.754 \pm 0.046$ \\
Digits & Amplitude  & \cellcolor{green!18}$0.324 \pm 0.046$ & \cellcolor{green!18}$0.314 \pm 0.040$ & \cellcolor{green!18}$0.325 \pm 0.046$ & $0.780 \pm 0.020$ \\
Digits & SPATE      & $0.307 \pm 0.030$ & $0.253 \pm 0.047$ & $0.307 \pm 0.029$ & \cellcolor{green!18}$0.799 \pm 0.019$ \\
\hline
Moons  & Angle      & $0.630 \pm 0.049$ & $0.643 \pm 0.057$ & $0.593 \pm 0.083$ & $0.721 \pm 0.082$ \\
Moons  & Amplitude  & $0.490 \pm 0.020$ & $0.394 \pm 0.198$ & $0.527 \pm 0.373$ & $0.446 \pm 0.039$ \\
Moons  & SPATE      & \cellcolor{green!18}$0.840 \pm 0.043$ & \cellcolor{green!18}$0.870 \pm 0.029$ & \cellcolor{green!18}$0.800 \pm 0.087$ & \cellcolor{green!18}$0.923 \pm 0.027$ \\
\hline
Circles & Angle     & \cellcolor{green!18}$0.797 \pm 0.050$ & \cellcolor{green!18}$0.753 \pm 0.060$ & \cellcolor{green!18}$0.893 \pm 0.044$ & \cellcolor{green!18}$0.879 \pm 0.083$ \\
Circles & Amplitude & $0.433 \pm 0.053$ & $0.427 \pm 0.059$ & $0.433 \pm 0.114$ & $0.399 \pm 0.062$ \\
Circles & SPATE     & $0.647 \pm 0.083$ & $0.682 \pm 0.125$ & $0.573 \pm 0.202$ & $0.753 \pm 0.081$ \\
\hline
Blobs & Angle       & $0.890 \pm 0.147$ & $0.893 \pm 0.144$ & $0.890 \pm 0.147$ & $0.956 \pm 0.080$ \\
Blobs & Amplitude   & $0.963 \pm 0.024$ & $0.966 \pm 0.021$ & $0.963 \pm 0.024$ & $0.998 \pm 0.002$ \\
Blobs & SPATE       & \cellcolor{green!18}$0.983 \pm 0.033$ & \cellcolor{green!18}$0.987 \pm 0.027$ & \cellcolor{green!18}$0.983 \pm 0.033$ & \cellcolor{green!18}$1.000 \pm 0.000$ \\
\hline
\end{tabular}%
}
\end{table}

To evaluate whether the representation improvements induced by SPATE translate into practical gains in a trainable hybrid QNN, Table~\ref{tab:qnn_results_summary} reports accuracy, precision, recall, and AUC under the same cross-validation protocol. The clearest gains appear on datasets where SPATE also improves representation geometry. On \textbf{Wine}, SPATE reaches $0.826\pm0.071$ accuracy and $0.978\pm0.018$ AUC, outperforming amplitude ($0.685\pm0.023$ accuracy, $0.887\pm0.026$ AUC) and angle ($0.399\pm0.063$ accuracy, $0.552\pm0.050$ AUC). On \textbf{Moons}, SPATE improves accuracy to $0.840\pm0.043$ and AUC to $0.923\pm0.027$, consistent with its stronger clustering in Figs.~\ref{fig:tsne_a}--\ref{fig:tsne_b}. On \textbf{Cancer}, SPATE improves accuracy to $0.837\pm0.062$ and achieves high recall ($0.983\pm0.011$), while angle/amplitude remain around $0.60$--$0.62$ accuracy.

To clarify when SPATE is less advantageous, two regimes are particularly informative. On \textbf{Circles}, angle remains best ($0.797\pm0.050$ accuracy, $0.879\pm0.083$ AUC), consistent with its cleaner separation pattern and stronger alignment scores. On \textbf{Digits}, accuracy remains similar across encodings ($\approx 0.307$--$0.324$), but SPATE yields the highest AUC ($0.799\pm0.019$), suggesting that in multi-class and capacity-limited settings, gains may appear primarily in ranking behavior rather than top-1 accuracy.

\subsection{Discussion and Key Takeaways}
\label{subsec:takeaways}
To summarize the main conclusions, SPATE generally improves representation geometry and often translates these improvements into better classification performance, while also exposing clear boundary cases where simpler encodings are better matched to the data geometry. The strongest evidence appears on Moons, Wine, Cancer, and Blobs, whereas Circles remains a counterexample, favoring angle, and Digits highlights limitations under strict multi-class capacity constraints.

\begin{itemize}
    \item SPATE produces more class-consistent and more separable embeddings on most datasets, with large gains in CKTA/Fisher/Silhouette in Table~\ref{tab:encoding_metrics_cv}.
    \item The t-SNE plots show clearer neighborhoods for SPATE in the same datasets where separability metrics improve (notably Moons and Blobs), supporting a geometric interpretation of the gains.
    \item Improvements in embedding geometry frequently translate into better hybrid-QNN performance, with the clearest gains on Wine, Moons, and Cancer in Table~\ref{tab:qnn_results_summary}.
    \item Dataset geometry matters: angle remains preferable for Circles, where periodic rotation mappings naturally fit ring-shaped boundaries.
    \item In capacity-limited multi-class settings (Digits), SPATE improves AUC and local organization, but accuracy remains constrained, suggesting that higher quantum capacity or more expressive readout may be required for larger-class regimes.
\end{itemize}
\section{Conclusion}
This work proposes SPATE, a spike-informed state preparation method that injects rate, spike timing phase, and coarse temporal structure into a shallow circuit using feature qubits and a compact time-qubit register. The results show that improving the embedding geometry at the encoding stage can yield more label-aligned and more separable representations under the same qubit budget, and these gains can translate into higher classification performance for a fixed hybrid QNN on several datasets. At the same time, the comparison highlights that no single encoding is universally best, and dataset geometry can favor simpler mappings in specific cases.

SPATE is not tied to a particular classifier and can be used as a general encoding prefix for QNNs and other QML pipelines, or in embedding mode when a probability-vector representation is needed. A key limitation is reliance on spike-generation and temporal-binning hyperparameters, which can affect performance across tasks; future work will investigate more systematic selection strategies and evaluate robustness to noise and hardware.

\section*{Acknowledgment}
This work was supported in part by the NYUAD Center for Quantum and Topological Systems (CQTS), funded by Tamkeen under the NYUAD Research Institute grant CG008, and the Center for Cyber Security (CCS), funded by Tamkeen under the NYUAD Research Institute Award G1104.

\bibliographystyle{IEEEtran}

\bibliography{refs}

\end{document}